\documentclass[aps,prd,twocolumn,nofootinbib,groupedaddress,amsfonts,floatfix]{revtex4}

\usepackage{graphicx,amsmath,amssymb,amstext}
\usepackage{amssymb,amsbsy,amsfonts,amsthm}
\usepackage{lscape}

\newcommand{\gsim}{\;\mbox{\raisebox{-0.5ex}{$\stackrel{>}{\scriptstyle{\sim}}$}}\;}

\usepackage{color}
\usepackage{ifthen}
\newboolean{editorial}
\setboolean{editorial}{true}
\newcommand{\editorial}[2]{\ifthenelse{\boolean{editorial}}{\textcolor{red}{[\textsf{\textbf{{#1}}}: }\textcolor{blue}{\textsf{{#2}}}\textcolor{red}{]}}{}}

\begin{document}

\title{Vacuum Bubbles in the Presence of a Relativistic Fluid}

\author{John T. Giblin, Jr${}^{1,2}$}
\author{James B. Mertens${}^2$}

\affiliation{${}^1$Department of Physics, Kenyon College, Gambier, OH 43022}
\affiliation{${}^2$CERCA/ISO, Department of Physics, Case Western Reserve University, Cleveland, OH}

\begin{abstract}
First order phase transitions are characterized by the nucleation and evolution of bubbles.  The dynamics of cosmological vacuum bubbles, where the order parameter is independent of other degrees of freedom, are well known; more realistic phase transitions in which the order parameter interacts with the other constituents of the Universe is in its infancy.  Here we present high-resolution lattice simulations that explore the dynamics of bubble evolution in which the order parameter is coupled to a relativistic fluid.  We use a generic, toy potential, that can mimic physics from the GUT scale to the electroweak scale.
\end{abstract}

\maketitle

\section{Introduction}
\label{intro}

The Universe had a hot early state--most commonly thought to be a result of post-inflationary reheating.  As the Universe cooled, over many orders of magnitude, phase transitions likely occurred at many different scales \cite{Kirzhnits:1972iw,Kirzhnits:1976ts}.  There were at least two associated with the Standard Model, the quantum-chromodynamic (QCD) and electro-weak phase transitions, but there may have also been others at higher energy scales.

The possibility of observing gravitational radiation from these phase transitions is exciting.  In the Standard Model, the Electroweak phase transition is second-order \cite{Kajantie:1996mn,Laine:1998vn} as the Higgs field smoothly transitions from its high-temperature symmetric state to its low-temperature vacuum expectation value (VEV).  However, only minimal modifications to the Higgs potential from new physics beyond the Standard Model \cite{Carena:1996wj,Delepine:1996vn,Laine:1998qk,Grojean:2004xa,Huber:2000mg,Huber:2006wf,Laine:2012jy,Carena:2008vj,Carena:2012np} can force this transition to be first-order.  In this case a tunneling process moves the Higgs field to the vacuum state in the potential at which time the Higgs acquires a VEV.  Bubbles form, collide, and coalesce, greatly increasing the possibility of an observable cosmological phase transition by means of searching for the gravitational wave background.

Simulations of vacuum bubbles at different scales confirm that we understand the evolution of these bubbles in many different models.  In the vast majority of these simulations, the field that undergoes the phase transition is uncoupled to other degrees of freedom and the velocities of the walls of the bubbles quickly approach the speed of light.  Early simulations reduced the number of dimensions of the problem \cite{Hawking:1982ga,Kosowsky:1991ua,Turner:1992tz,Kosowsky:1992rz,Kosowsky:1992vn} capturing the physics of the direct collisions of two bubbles.  The potentials that govern these fields have been growing more complicated, the number of dimensions in the simulations have grown, and much has been learned about the rich dynamics systems of bubbles colliding  \cite{Giblin:2010bd,Easther:2009ft,Kosowsky:2001xp,Gogoberidze:2007an,Caprini:2006jb,Caprini:2009yp}.  These advancements have allowed authors to couple the fundamental degree of freedom to other scalar fields, and have shown how energy can be exchanged between the two.

The challenge at the electro-weak scale is greater.  Fermions and gauge-bosons coupled to the Higgs field do not have a semi-classical interpretation and cannot be simulated alongside the Higgs field on a classical lattice.  More importantly, the dynamics of the coupled fields are not of the same scale as the dynamics of the Higgs field.  During the electro-weak phase transition it is more appropriate to think of the Standard Model fields as a coupled, relativistic fluid. 

There has been recent work in this area.  A new study of a toy model in 1+1-dimensions \cite{Huber:2013kj} has started to probe this question, studying the numerical evolution of dimensionally-reduced bubbles in the presence of a fluid.  A subset of these authors, joined by others, have followed up with a a 3+1-dimensional simulation for a specific set of parameters that yields a first-guess at the gravitational radiation produced by the field during this process \cite{Hindmarsh:2013xza}, although this work does not directly closely examine the dynamics of the bubbles themselves in the coupled field/fluid system.  We believe that explicit investigation of the numerical system is necessary first. 

Therefore before we direct our attention to the the generation of gravitational radiation, we have to fully understand the evolution of bubbles in such a system.  Here we will parameterize the phase transition with a scalar field subject to a canonical kinetic term,
\begin{equation}
S_\phi = \int\mathrm{d}^{4}x \left(-\frac{1}{2}\partial_{\mu}\phi\partial^{\mu}\phi - V\left(\phi\right)  \right)
\label{faction}
\end{equation}
where, for now, we do not specify a coupling between the field and fluid. In order to broadly talk about first-order phase transitions of different strengths, we parameterize the potential $V(\phi)$ as in \cite{Dunne:2005rt} using
\begin{equation}
V(\phi) = \frac{m^2}{2} \phi^2 + \eta \phi^3 + \frac{\lambda}{8} \phi^4.
\label{unscaledpot}
\end{equation}
This choice of potential gives us flexibility, as it allows for comparisons between the standard thin-wall parameterization of first-order phase transitions, e.g. \cite{Kosowsky:1991ua,Kosowsky:1992vn,Child:2012qg}, as well as models in which the temperature dependence of the potential is explicitly included \cite{Huber:2013kj,Kamionkowski:1993fg,Hindmarsh:2013xza}.  Note that we have changed the sign of the field (and therefore the cubic term of Eq.~\ref{unscaledpot}) relative to \cite{Dunne:2005rt} so that the true vacuum lies in the region where $\phi<0$.

\begin{figure}
  \centering
    \includegraphics[width=0.45\textwidth]{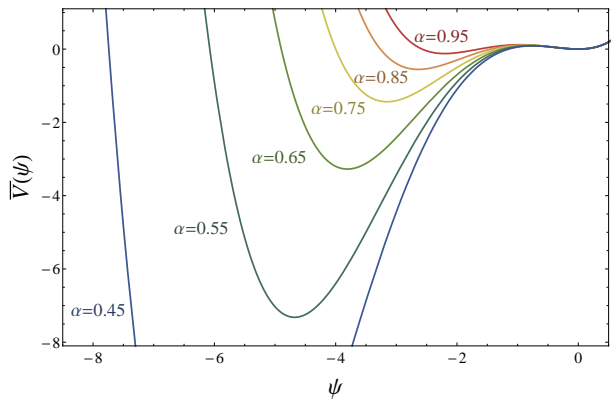}
  \caption{\label{potfig}The unit-less potential, $\bar{V}(\psi)$, for different values of $\alpha$.  We cut off the $\alpha=0.45$ curve to retain some details of the curves for higher values of $\alpha$.}
\end{figure}

We can further define a dimensionless field, 
\begin{equation}
\psi = \frac{2\eta}{m^2} \phi
\end{equation}
and dimensionless coordinates,
\begin{equation}
\bar{x}^\mu = m\, x^\mu,
\end{equation}
which we can use to convert our action, Eq.~\ref{faction}, to
\begin{equation}
S_\phi = \left(\frac{m^2}{4\eta^2}\right)\int d^4\bar{x} \, \left(- \frac{1}{2}\bar{\partial}^\mu\psi \bar{\partial}_\mu\psi - \bar{V}(\psi)\right).
\end{equation}
with an effective, dimensionless potential,
\begin{equation}
\label{alphapot}
\bar{V}(\psi) = \frac{1}{2}\psi^2 + \frac{1}{2}\psi^3 + \frac{\alpha}{8}\psi^4
\end{equation}
parameterized by
\begin{equation}
\alpha = \frac{\lambda m^2}{4\eta^2}.
\end{equation}
This construction defines a broad class of possible first-order phase transitions.  Fig.~\ref{potfig} shows how Eq.~\ref{alphapot} can realize phase transitions in which the two minima are approximately the same, $\alpha \rightarrow 1$, which provides comparison with historical work on phase transitions in cosmological context (e.g., \cite{Child:2012qg,Kamionkowski:1993fg}) as well as those more common in the particle physics literature $\alpha \sim 0.5$ (e.g., \cite{Ignatius:1993qn}).

In this work we will examine three scenarios, (I) The thin-wall limit, where $\alpha = 0.96$, (II)  $\alpha = 0.65$ which is used in \cite{KurkiSuonio:1995vy}  and very similar to the value $\alpha=0.76$ \cite{Ignatius:1993qn} associated with a historically interesting standard model electro-weak phase transition, and (III) $\alpha = 0.45$ a parameterization in which the bubble walls are about as thick as the radii of the bubbles themselves.  This is described as an `intermediate strength transition' in \cite{Hindmarsh:2013xza}.  

In the thin-wall limit, we are used to seeing an almost-degenerate potential parameterized by 
\begin{equation}
V(\phi) = \frac{\lambda_{TW}}{8}\left(\phi^2-\phi_0^2\right)^2 +\epsilon \lambda_{TW} \phi_0^3 (\phi+ \phi_0)
\end{equation}
where one can translate between the two with\begin{equation}
\alpha \approx 1 - \epsilon + \mathcal{O}(\epsilon^2).
\end{equation}

Computationally, case I is the most difficult, as the bubbles produced in this model have thin walls (as discussed in more detail below) and require higher resolution in order to evolve.

Since we will evolve our bubbles on a fixed, non-expanding grid, the vacuum energy of the true vacuum does not enter into our equations of motion, and thus we choose not to specify it.  The overall energy density of the simulation would be important if we included gravitational effects or if we tied the simulation to present-day observables; we leave these two extensions to future work.  Here we are more interested in the generic behavior of phase transitions and therefore do not constrain our simulations to a particular energy scale.

The relative contributions of the fluid and field to the overall energy density of the Universe are important.  We can define the ratio between the vacuum energy and fluid energy density to be
\begin{equation}
\beta = \frac{\left(\sqrt{9-8 \alpha }+3\right)^2 \left(-4 \alpha +\sqrt{9-8 \alpha }+3\right)}{64 \alpha^3 \bar{\epsilon}_{fl}}.
\end{equation}
For a well-motivated phase transition, the parameter $\beta$ should be known.  Here we will keep it as a free parameter, understanding that the $\beta \rightarrow 0$ and $\beta \rightarrow \infty$ limits make the physics of the fluid and field dominant respectively.

We organize this paper as follows.  In Section~\ref{bubnuc} we will describe how we set the initial conditions for bubbles in our three scenarios.  We will then introduce our dynamical systems for both the scalar field and coupled relativistic fluid in section~\ref{fluiddynamics}.  We present the results of our simulations in section~\ref{results} and conclude in section~\ref{discussion}.

\section{Bubble Nucleation}
\label{bubnuc}

A first order phase transition progresses as regions of space, bubbles, tunnel from the meta-stable configuration of the false vacuum to the the true vacuum.  These regions eventually collide and coalesce until the entire volume is in the true vacuum.  We must begin by understanding the tunneling event(s) between the two minima.  Assuming that the bubble is maximally symmetric, this solution, also known as the classical ``bounce'' solution \cite{Coleman:1977py,Coleman:1980aw}, is a solution to the differential equation
\begin{equation}
\label{4deq}
  \psi''(\bar{\rho}) + \frac{3}{\bar{\rho}}\psi'(\bar{\rho}) = \bar{V}'(\psi)
\end{equation}
with boundary conditions
\begin{eqnarray}
  \psi'(0) & = 0 \label{eq:4deqcnstrns1} \\
  \psi(\bar{\rho} \rightarrow \infty) & = 0 . \label{eq:4deqcnstrns2}
\end{eqnarray}
Traditionally the approximate--hyperbolic tangent--solution to this equation, as in \cite{Coleman:1977py,Coleman:1980aw}, is sufficient for analytic and numerical work.  Here, however, we go beyond this approximation since we depart significantly from the thin-wall limit.  As such, we employ a numerical scheme, the shooting method, similar to that of \cite{Dunne:2005rt} to create our initial conditions.

\begin{figure}
  \centering
    \includegraphics[width=0.45\textwidth]{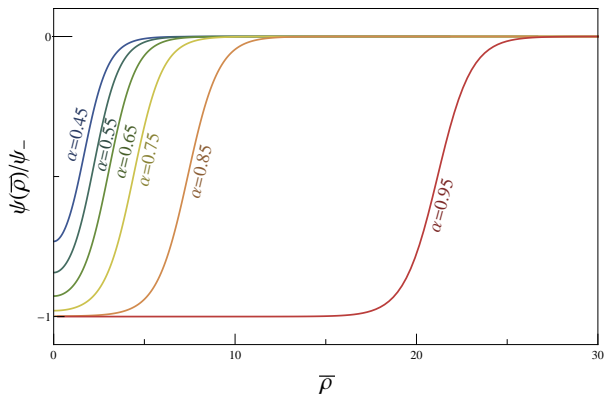}
  \caption{\label{bubfig}The instanton profiles for different values of $\alpha$.  Field values are rescaled by the difference in field values in the minima, equivalent to the field values in the true vacuum, $\psi_{-}$.}
\end{figure}

To do this, we discretize Eq. \ref{4deq} on a 1-dimensional grid, which extends out a large distance $\bar{\rho}$ from the expected location of the bubble wall $\bar{R}_0$.  We begin by setting the field value furthest from the bubble equal to the field value in the false vacuum (effectively applying Eq.~\ref{eq:4deqcnstrns2}).  We then ``undershoot" by guessing a much lower field value in the adjacent grid cell, and solve for successive values of the field using a discretized version of Eq. \ref{4deq}.  We then raise the value of the guess until we ``overshoot", a condition determined by a sign change of the derivative in the bubble interior.  We continue refining the guess until Eq. \ref{eq:4deqcnstrns1} is satisfied to within an arbitrary tolerance.  Fig.~\ref{bubfig} shows the results of this analysis for varying values of $\alpha$.  When $\alpha\approx 1$, the thickness of the wall is (much) smaller than the scale of the bubble radius \cite{Dunne:2005rt},
\begin{equation}
\bar{R_0} = m R_0 \approx (1-\alpha)^{-1}.
\end{equation}
In this case, the interior of the bubble is stable in the lower minima.  This is expected when we are close to the thin-wall limit.  However, as $\alpha$ decreases, the bubble radius becomes smaller than the thickness of the wall and the instanton solution no longer interpolates all the way to the lower minima.  In this regime the bubbles have a non-analytic profile, and the entire interior region is dynamical as the bubble evolves.

\section{Relativistic Fluids}
\label{fluiddynamics}

The action for a relativistic fluid can be written as \cite{Schutz:1970my}
\begin{equation}
S = \left(\frac{m^2}{4\eta^2}\right)\int\mathrm{d^{4}}\bar{x}\sqrt{-\bar{g}}\left(\frac{\bar{R}}{2\bar{\kappa}}+\mathcal{\bar{L}}_{fluid}\right),
\end{equation}
where the fluid Lagrangian density is the dimensionless pressure,
\begin{equation}
\mathcal{\bar{L}}_{fluid} = \bar{p} = \frac{4\eta^2}{m^6}p, 
\end{equation}
which is parameterized by the dimensionless fluid energy density $\bar{\epsilon}$ and fluid velocity $U^\mu$.  In this language, the Ricci scalar and Planck mass have dimensionless analogues,
\begin{equation}
\bar{R} = R/m^2,\ \ \ \bar{\kappa} = \frac{m^4}{4\eta^2}\kappa.
\end{equation}
Since the fluid is relativistic, it satisfies $U^\mu U_\mu = -1$.  To derive equations of motion for the dynamical components of the model we follow a traditional method, e.g. \cite{Schutz:1970my}.  We start by identifying the dimensionless stress-energy tensor is
\begin{eqnarray}
\bar{T}^{\mu\nu} & = & \frac{-2}{\sqrt{-\bar{g}}}\frac{\delta\sqrt{-\bar{g}}\bar{p}}{\delta \bar{g}^{\mu\nu}}\\
 & = & (\bar{\epsilon}+\bar{p})U^{\mu}U^{\nu}+\bar{p}\bar{g}^{\mu\nu}.
\end{eqnarray}
Which, in the absence of coupling, is exactly conserved, $\bar{D}_{\mu}\bar{T}^{\mu\nu} = 0 $.  We use this conservation to construct evolution equations for the components of the four-velocity, $U^\mu$.  For the time being, we will assume a general equation of state for the fluid, $\bar{p} = w \bar{\epsilon}$, and allow for a generic source $j^\nu$,
\begin{equation}
\label{evolution1}
\bar{D}_{\mu}\bar{T}^{\mu\nu} = (1+w)\bar{D}_{\mu}(\bar{\epsilon} U^{\mu}U^{\nu})+w\bar{D}^{\nu}\bar{\epsilon} = j^\nu.
\end{equation}

One can then contract Eq.~\ref{evolution1} with $U^\nu$ to produce a continuity equation for the fluid energy density,
\begin{equation}
\label{conserve}
\bar{D}_{\mu}(\bar{\epsilon} U^{\mu})=\left(\frac{1}{1+w}\right)\left(w U^{\mu}\bar{D}_{\mu}\bar{\epsilon}-U_{\mu}j^{\mu}\right).
\end{equation}
For a non-relativistic fluid consisting of ordinary matter in flat space ($U^{i} = \gamma\vec{v}\simeq\vec{v}$, $w = 0$,  $D_{\mu} = \partial_{\mu}$, and $U_\mu j^\mu = \sigma$), Eq.~\ref{conserve} reduces to the usual Navier-Stokes equation,
\begin{equation}
\frac{\partial\rho}{\partial t} + \vec{\nabla}\cdot(\rho\vec{v}) = \sigma.
\end{equation}

The next step is to find equations of motion for the $U^\mu$ and $\epsilon$.  We note that we only need to calculate the spatial part of the four-velocity, $U^i$, as normalization requires, $U^t = (1 + U^i U_i)^{1/2}$.  Eq.~\ref{conserve} eventually leads us to the two equations,
\begin{equation}
\label{spaceeom}
\begin{split}
\bar{\partial}_{t}U^{i} =& \frac{wU^{t}U^{i}}{\left(1+(1-w)U^{k}U_{k}\right)} \\
& \times \left( \bar{\partial}_{k}U^{k}-\frac{U_{j}}{\left(U^{t}\right)^{2}}\left(U^{k}\bar{\partial}_{k}U^{j}+\frac{w}{1+w}\bar{\partial}^{j}\ln\left(\bar{\epsilon}\right)\right)\right) \\
& - \frac{1}{U^{t}}\left(U^{k}\bar{\partial}_{k}U^{i}+\frac{w}{1+w}\bar{\partial}^{i}\ln\left(\bar{\epsilon}\right)\right)+J^{i}
\end{split}
\end{equation}
and
\begin{equation}
\label{energyeom}
\begin{split}
\bar{\partial}_{t}\ln\left(\bar{\epsilon}\right) =& -\frac{(1+w)U^{t}}{\left(1+(1-w)U^{k}U_{k}\right)} \\
& \times \left(\frac{1-w}{1+w}U^{i}\bar{\partial}_{i}\ln\left(\bar{\epsilon}\right)+\bar{\partial}_{i}U^{i}+\frac{U_{j}U^{k}}{\left(U^{t}\right)^{2}}\bar{\partial}_{k}U^{j}\right) \\
& -\left(1+w\right)\frac{U^{i}J_{i}}{\left(U^{t}\right)^{2}} - \frac{1}{U^{t}}U^{\mu}\frac{j_{\mu}}{e^{\ln\left(\bar{\epsilon}\right)}}
\end{split}
\end{equation}
where
\begin{equation}
\begin{split}
J^{i} =& \frac{1}{e^{\ln\left(\bar{\epsilon}\right)} U^{t}}\left(\delta_{j}^{i}+\frac{w}{\left(1+(1-w)U^{k}U_{k}\right)}U^{i}U_{j}\right) \\
& \times \left(\frac{j^{j}}{1+w}+U^{j}U^{\mu}j_{\mu}\right),
\end{split}
\end{equation}
and which we numerically integrate to evolve the fluid.

As a further sanity check, we can look at the $1+1$ dimensional case in the absence of a coupling term.  In this case, $U^i = \{u^x\} = u$, and the evolution equations simply become,
\begin{eqnarray}
    \partial_{t}u & = & -\frac{u}{\sqrt{1+u^{2}}}\partial_{x}u\\
    \partial_{t}\epsilon & = & -\frac{1}{\sqrt{1+u^{2}}}\left(u\partial_{x}\epsilon+\epsilon\left(\frac{1}{1+u^{2}}\right)\partial_{x}u\right).
\end{eqnarray}
In the small $u$ limit, the first equation is recognizable as the inviscid Burgers' equation, and the second as a continuity equation.

It is worth pointing out that this treatment of a relativistic fluid is different from that proposed in \cite{KurkiSuonio:1995vy,Hindmarsh:2013xza}.  In these references, the authors use the cosmic-fluid-order-parameter-field model \cite{Ignatius:1993qn} where the potential 
\begin{equation}
\label{IKKLeq}
V_{KS}(\phi,T) = \frac{1}{2}\gamma \left(T^2-T_0^2\right)\phi^2 - \frac{1}{3}\alpha_{KS}T\phi^3+\frac{1}{4}\lambda_{KS}\phi^4
\end{equation}
includes explicit temperature dependence.  Note that we have included the subscript $KS$ to distinguish the parameters of Eq.~\ref{IKKLeq} from the parameters of Eqs.~\ref{unscaledpot} and \ref{alphapot}.  These authors then chose variables $Z_{KS}^i = \gamma (\epsilon_{KS} + p_{KS}) U^i$ and $E_{KS} = \gamma \epsilon_{KS}$, where
\begin{equation}
\epsilon_{KS} = 3 a T^4 + V_{KS} - T \frac{\partial V_{KS}}{\partial T}
\end{equation}
and
\begin{equation}
p_{KS} = a T^4 - V_{KS}.
\end{equation}
They use a set of evolution equations (Eqs.~9 and 10 in \cite{KurkiSuonio:1995vy} or Eqs.~6 and 7 in \cite{Hindmarsh:2013xza}), 
\begin{equation}
\label{eeomhh1}
\begin{split}
\dot{E}_{KS} + \partial_i(E_{KS} V^i) + p_{KS}[\dot{\gamma} + \partial_i(\gamma V^i)] \\
- \frac{\partial V_{KS}}{\partial \phi}\gamma (\dot{\phi} + V^i \partial_i \phi) 
  = \eta \gamma^2 (\dot{\phi} + V^i\partial_i\phi)^2
\end{split}
\end{equation}
and
\begin{equation}
\label{eeomhh2}
\begin{split}
\dot{Z}_{ES\,i} + \partial_j (Z_{ES\,i} V^j) + \partial_i p + \frac{\partial V_{KS}}{\partial \phi}\partial_i \phi \\
  = - \eta \gamma (\dot{\phi} + V^j \partial_j \phi) \partial_i \phi .
  \end{split}
\end{equation}
This set of equations, although more aesthetically pleasant than Eqs.~\ref{spaceeom} and \ref{energyeom}, is a more difficult system to evolve.  In particular, the $\dot{\gamma}$ term of Eq.~\ref{eeomhh1} requires knowledge of the time-differentiated components of the fluid three-velocity, which are not known quantities in the dynamical system.  The system defined by Eqs.~\ref{spaceeom} and \ref{energyeom} is closed.

Our choice of a temperature independent potential, Eq.~\ref{alphapot}, specifically neglects any spatial gradients in the temperature field that could be present during the phase transition.  The effects of temperature dependence seem negligible for the choices of parameters that we study.  Additionally it neglects the overall change in temperature of the Universe during the transition.  This last effect would, of course, be dominated by the global change in temperature due to the expansion of the Universe which all studies, including ours, neglect.

\subsection{Vacuum Bubbles in the presence of a fluid}

To set up a phenomenological model for the interaction between the fluid and field, we closely follow the cosmic-fluid-order-parameter-field model, as in \cite{KurkiSuonio:1995vy,Ignatius:1993qn,Hindmarsh:2013xza}.  The total stress-energy for the system is
\begin{equation}
\begin{split}
\bar{T}^{\mu\nu}_{fluid+\psi} =& \bar{\partial}^\mu \psi \bar{\partial}^\nu \psi - \bar{g}^{\mu\nu}\left( \frac{1}{2}\bar{\partial}_\alpha\psi\bar{\partial}^\alpha\psi + \bar{V}(\psi) \right) \\
& + (\bar{\epsilon} + \bar{p})U^\mu U^\nu + \bar{p} \bar{g}^{\mu\nu}
\end{split}
\end{equation}
which, in the absence of a coupling, should be conserved independently for the fluid and the field,
\begin{equation}
\bar{\partial}_\mu \bar{T}^{\mu\nu} = \bar{\partial}_\mu \bar{T}^{\mu\nu}_{\rm field} +  \bar{\partial}_\mu \bar{T}^{\mu\nu}_{\rm fluid} = 0+0.
\end{equation}
In the absence of a fundamental description of the interaction, we rely on a phenomenological interaction
\begin{equation}
j^\nu = \xi U^\mu \bar{\partial}_\mu\psi \bar{\partial}^\nu \psi,
\label{source}
\end{equation}
which we introduce to the two conservation equations.  The equation for the field,
\begin{equation}
\bar{\partial}_\mu \bar{T}^{\mu\nu}_{\rm field} = \bar{\Box} \psi \bar{\partial}^\nu \psi - \frac{\partial \bar{V}}{\partial\psi}\bar{\partial}^\nu \psi = -j^\nu,
\end{equation}
gives us the Klein-Gordon equation,
\begin{equation}
\bar{\Box} \psi = \frac{\partial \bar{V}}{\partial \psi} + \xi U^\mu \bar{\partial}_\mu \psi.
\end{equation}
The equation for the fluid,
\begin{equation}
\bar{\partial}_\mu \bar{T}^{\mu\nu}_{\rm fluid} = j^\nu
\end{equation}
is exactly Eq.~\ref{evolution1} with Eq.~\ref{source} identified as the source.

\section{Results}
\label{results}

We can now explore the three cases we identified in Section~\ref{intro}: (I) $\alpha = 0.96$, (II)  $\alpha = 0.65$ and (III)  $\alpha = 0.45$.  For different values of $\alpha$ we need different numerical parameters since the relative scales of the bubble wall thickness and bubble radius varies significantly from case to case.  Further, for each case, we chose a set of couplings, $\xi$, that show how the bubble evolution depends on the strength of the coupling.  
\begin{figure*}[p]
  \begin{tabular}{rccc}
    & $\bar{R} = \bar{R}_0$ & $\bar{R} = 1.5 \bar{R}_0$ & $\bar{R} = 2 \bar{R}_0$ \\
\\
    \rotatebox{90}{\hspace{5 mm} (a)~~$\xi = {\color[rgb]{0.246296, 0.315957, 0.80044} 0.0},~{\color[rgb]{0.324106, 0.60897, 0.708341} 0.0125},~{\color[rgb]{0.513417, 0.72992, 0.440682} 0.025},~{\color[rgb]{0.764712, 0.728302, 0.273608} 0.05},~{\color[rgb]{0.901627, 0.539872, 0.208366} 0.1},~{\color[rgb]{0.857359, 0.131106, 0.132128} 0.2}$}
    \hspace{5 mm}
      & \includegraphics[width=0.29\textwidth]{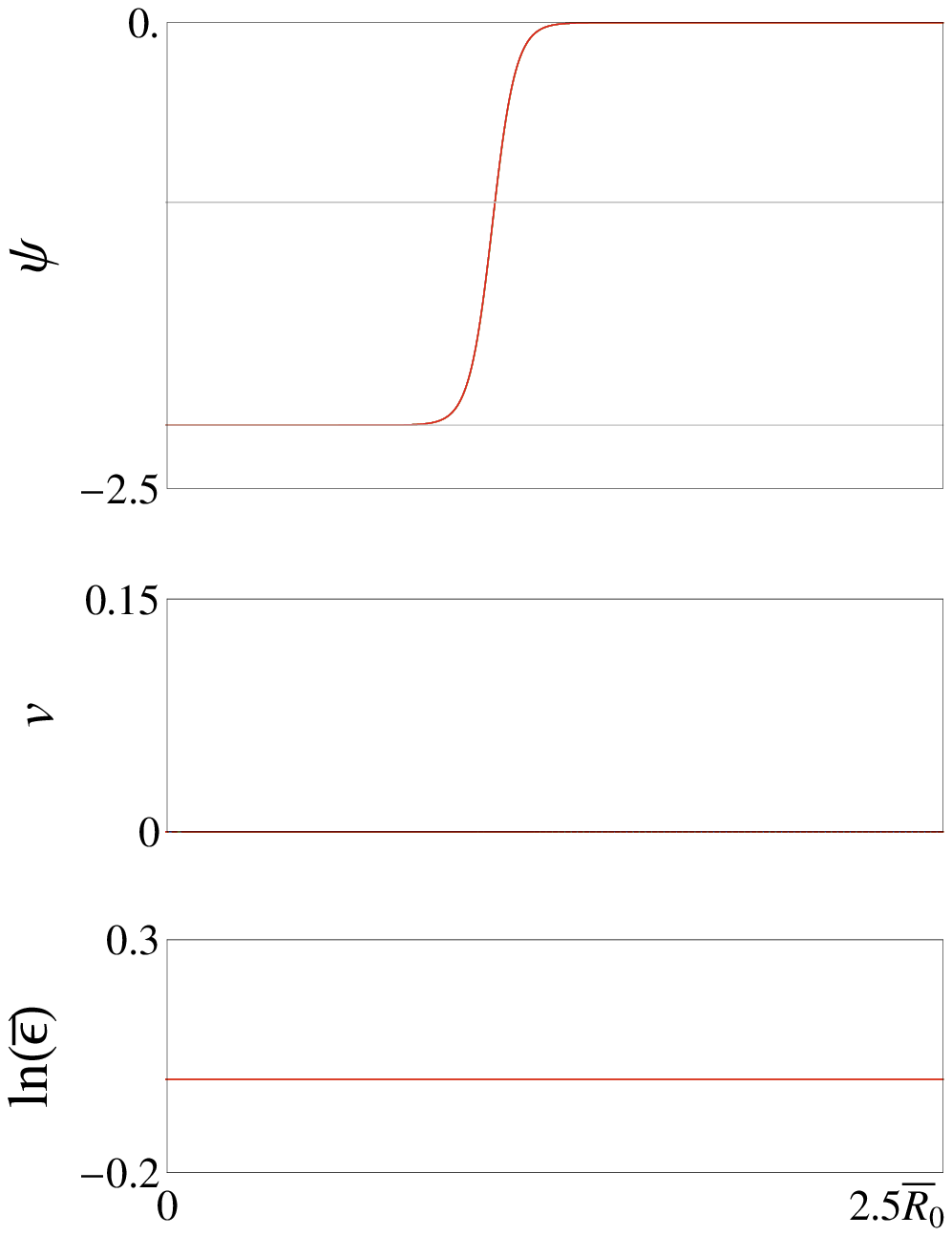}
      & \includegraphics[width=0.29\textwidth]{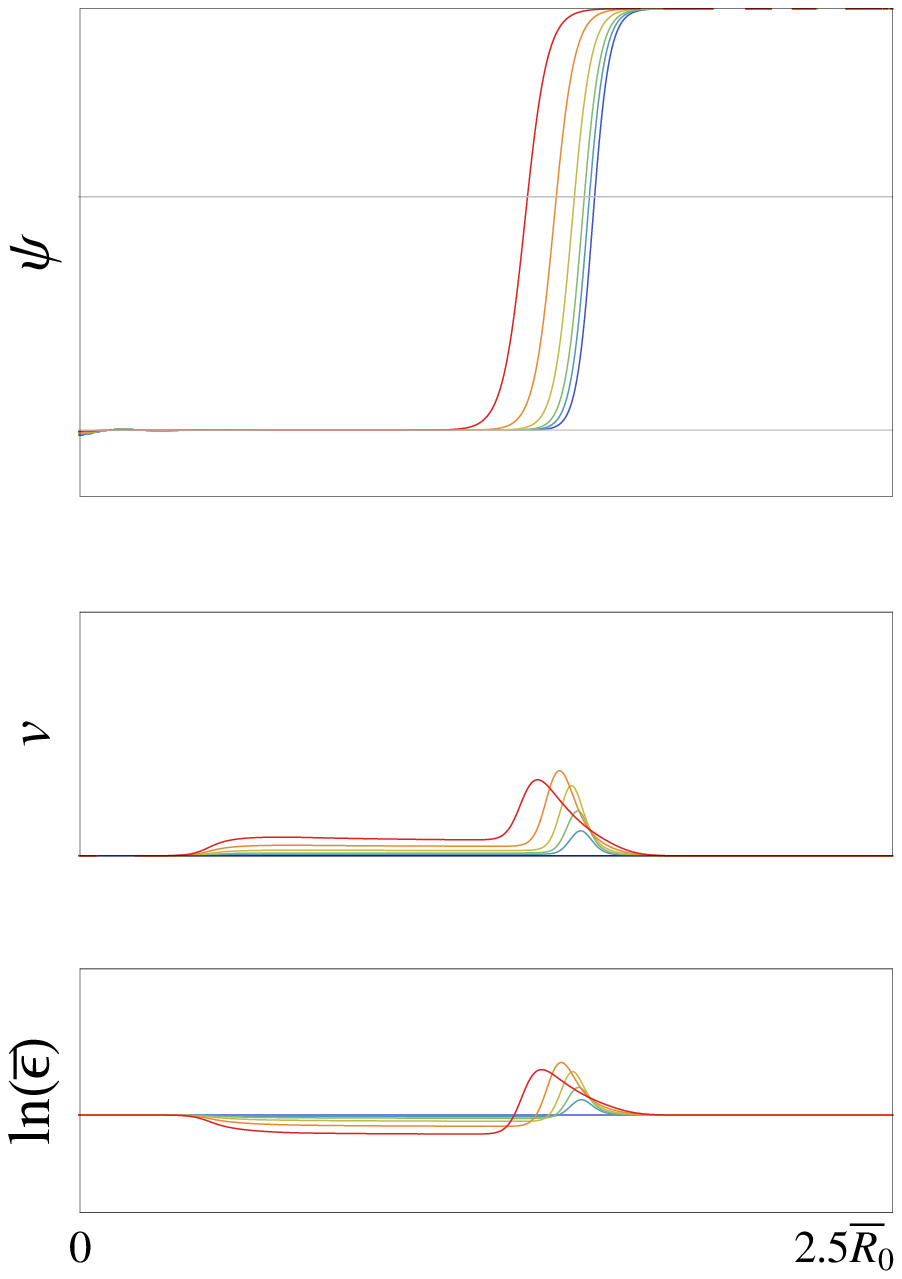}
      & \includegraphics[width=0.29\textwidth]{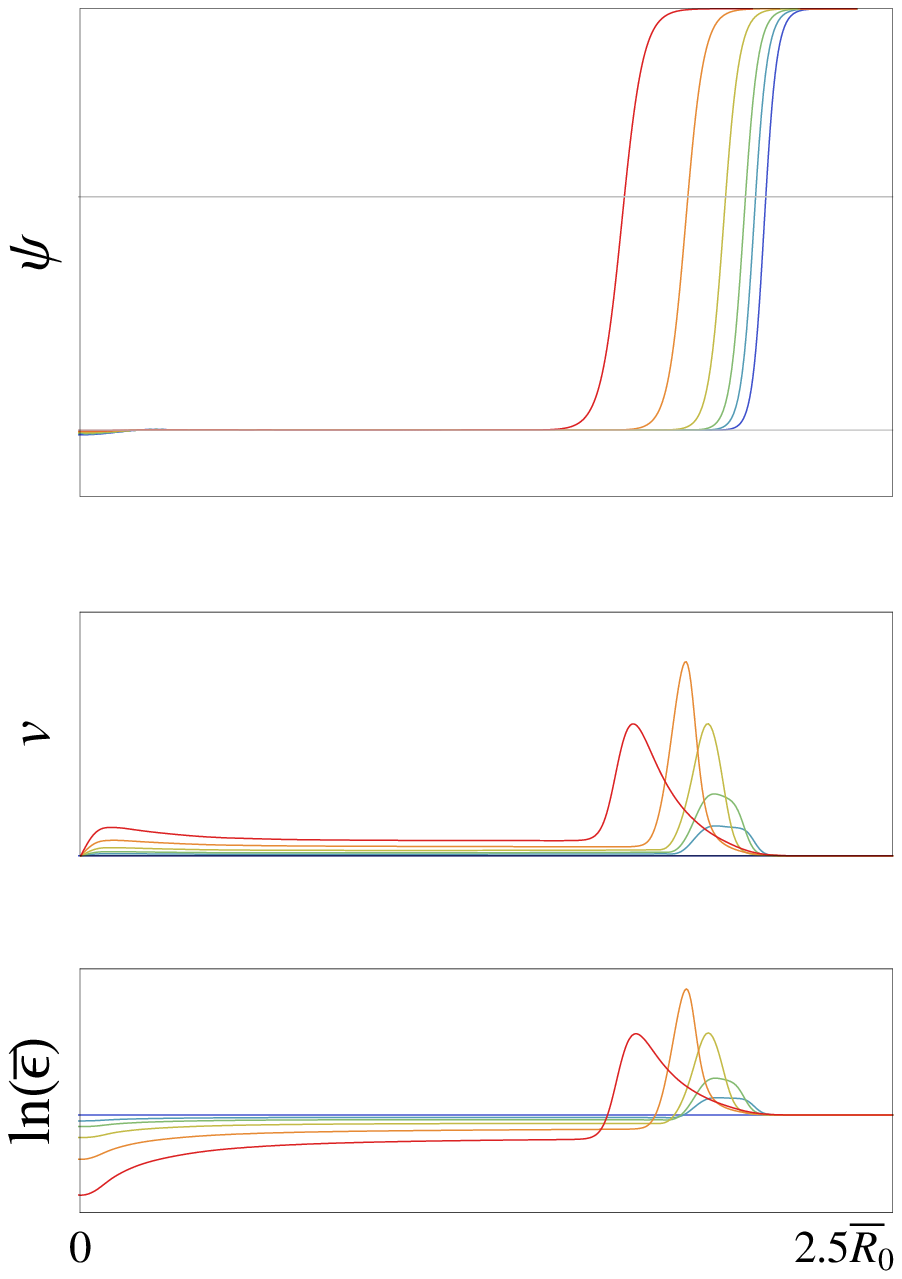} \\
\\ \\ 
    \rotatebox{90}{\hspace{5 mm} (b)~~$\xi = {\color[rgb]{0.246296, 0.315957, 0.80044} 0},~{\color[rgb]{0.324106, 0.60897, 0.708341} 0.0625},~{\color[rgb]{0.513417, 0.72992, 0.440682} 0.125},~{\color[rgb]{0.764712, 0.728302, 0.273608} 0.25},~{\color[rgb]{0.901627, 0.539872, 0.208366} 0.5},~{\color[rgb]{0.857359, 0.131106, 0.132128} 1.0}$}
    \hspace{5 mm}
      & \includegraphics[width=0.29\textwidth]{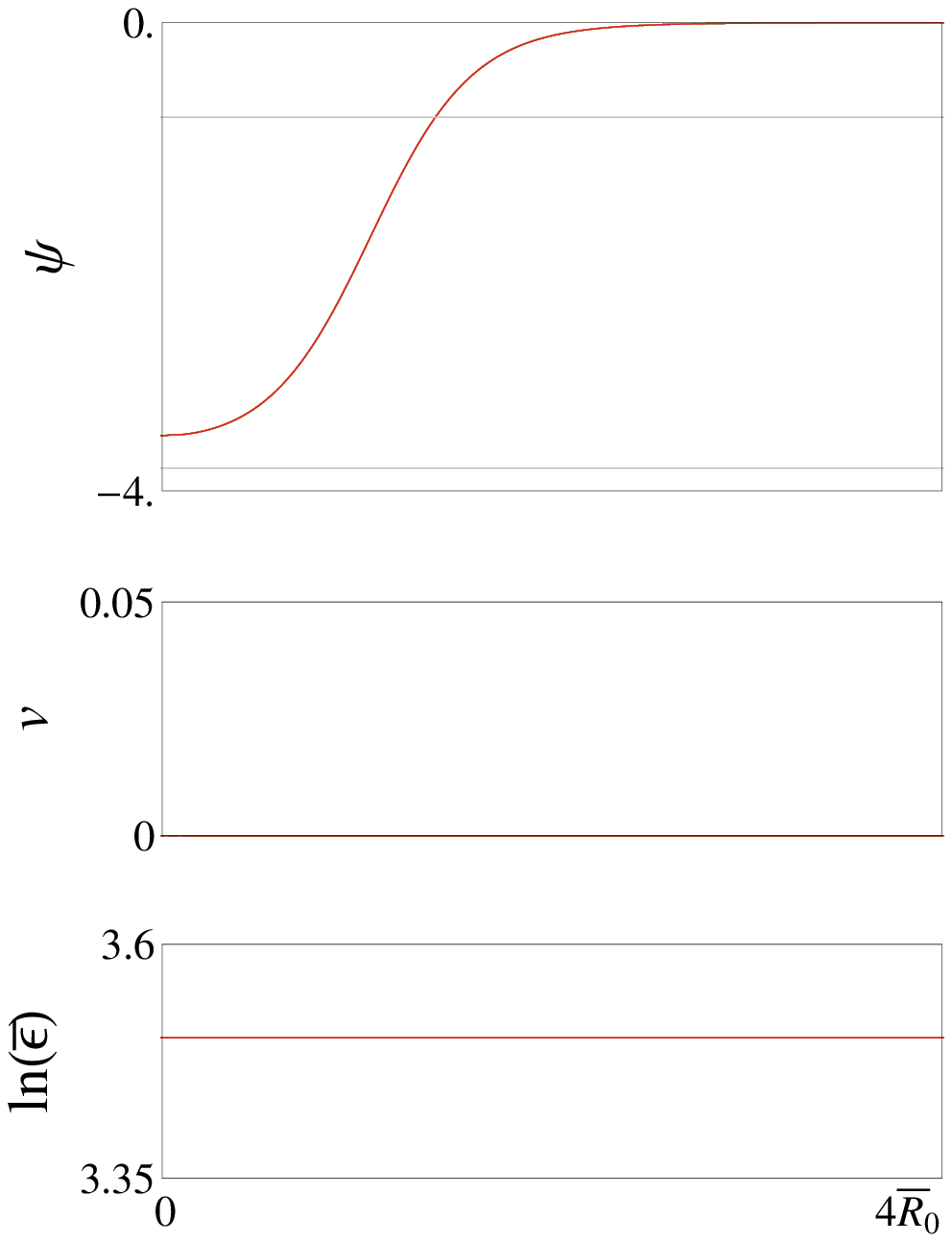}
      & \includegraphics[width=0.29\textwidth]{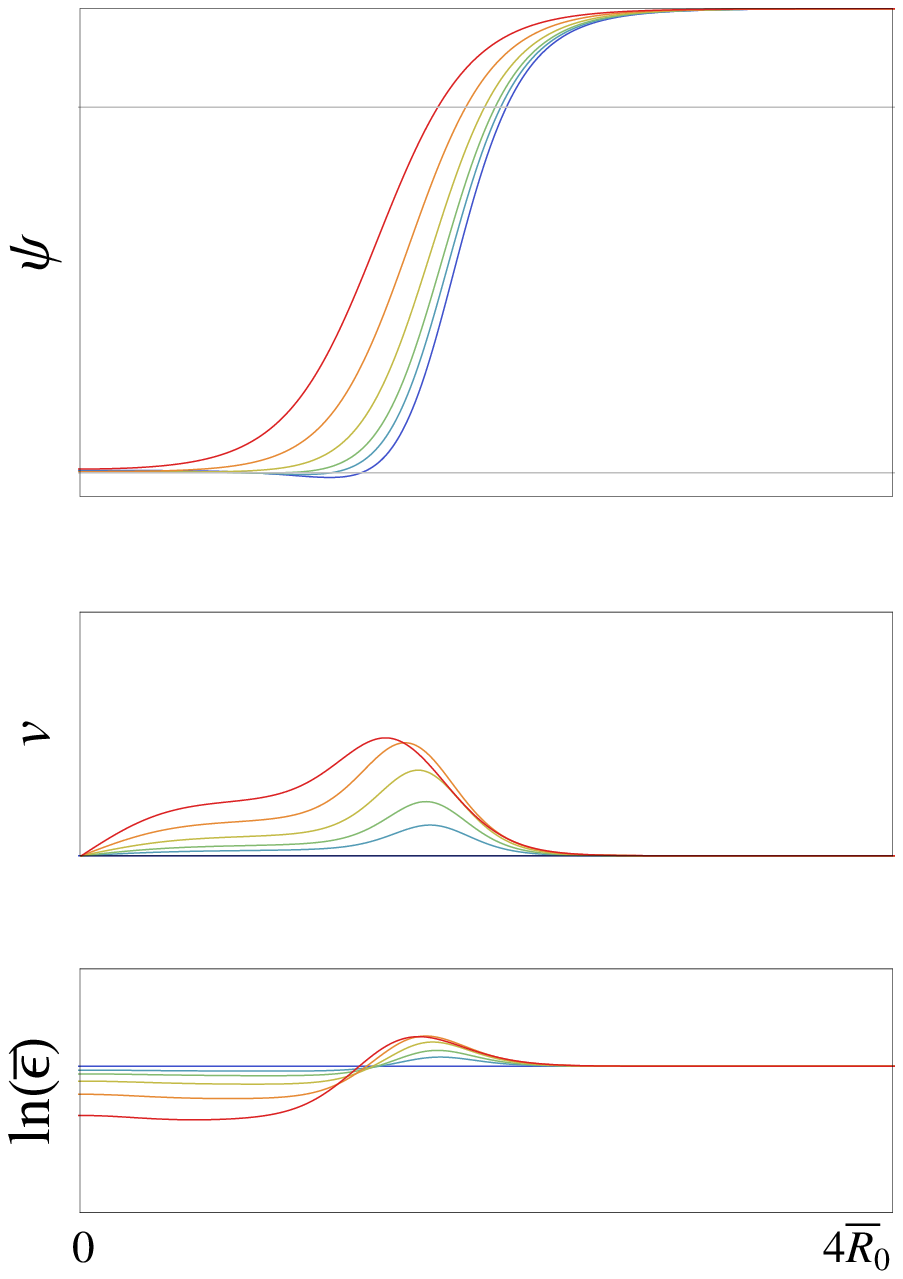}
      & \includegraphics[width=0.29\textwidth]{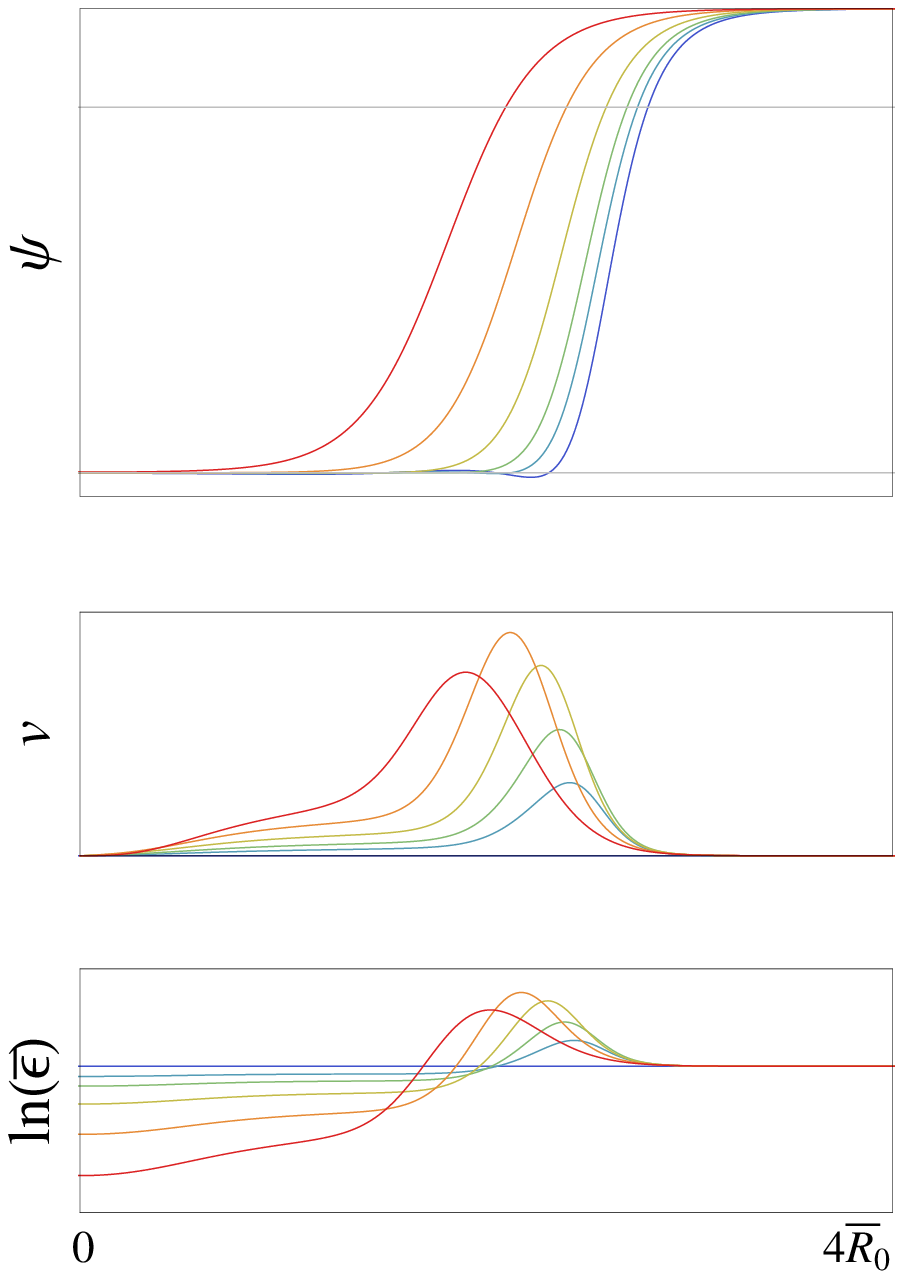} \\
\\ \\ 
   \rotatebox{90}{\hspace{5 mm} (c)~~~$\xi = {\color[rgb]{0.246296, 0.315957, 0.80044} 0},~{\color[rgb]{0.324106, 0.60897, 0.708341} 0.125},~{\color[rgb]{0.513417, 0.72992, 0.440682} 0.25},~{\color[rgb]{0.764712, 0.728302, 0.273608} 0.5},~{\color[rgb]{0.901627, 0.539872, 0.208366} 1.0},~{\color[rgb]{0.857359, 0.131106, 0.132128} 2.0}$}
   \hspace{5 mm}
      & \includegraphics[width=0.29\textwidth]{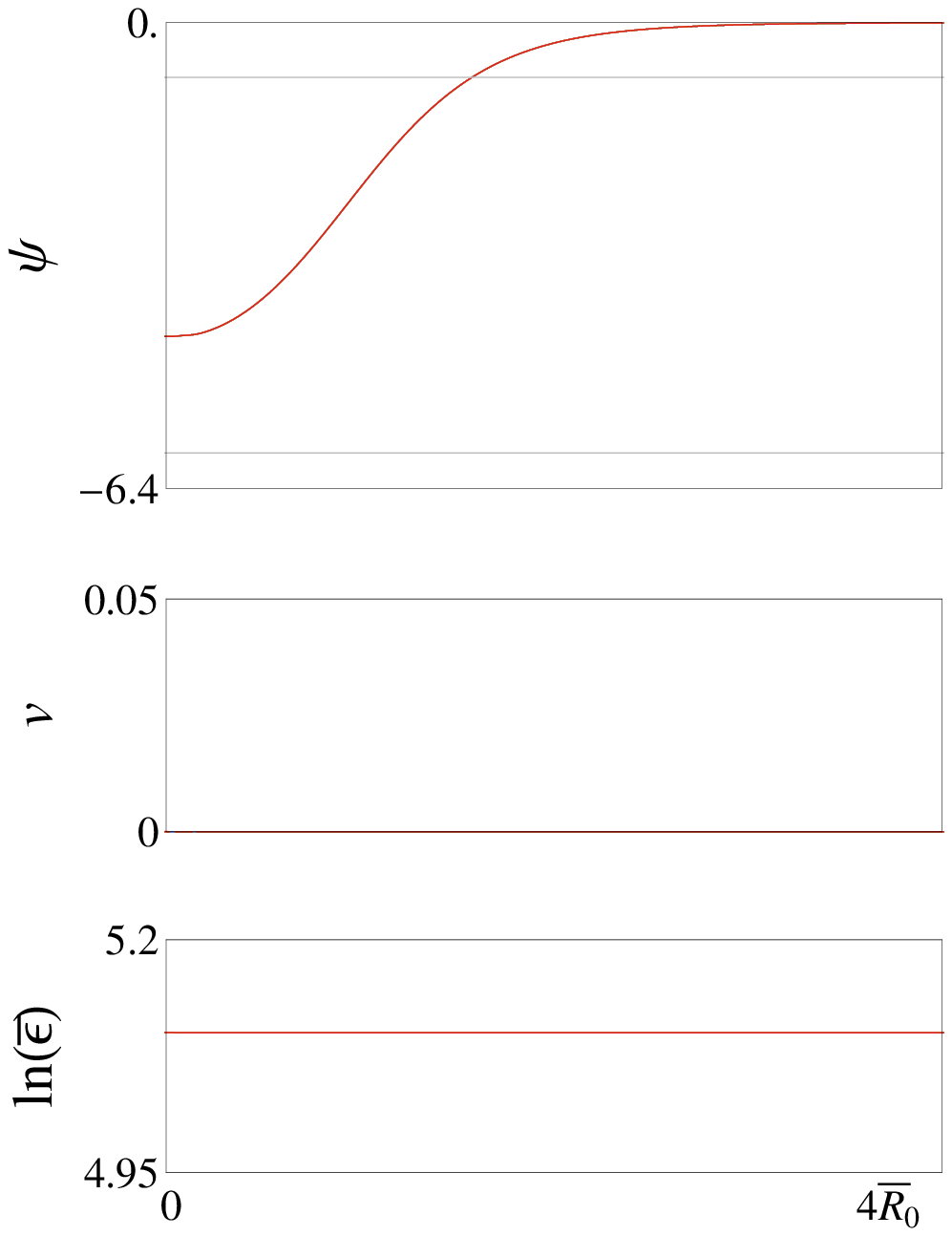}
      & \includegraphics[width=0.29\textwidth]{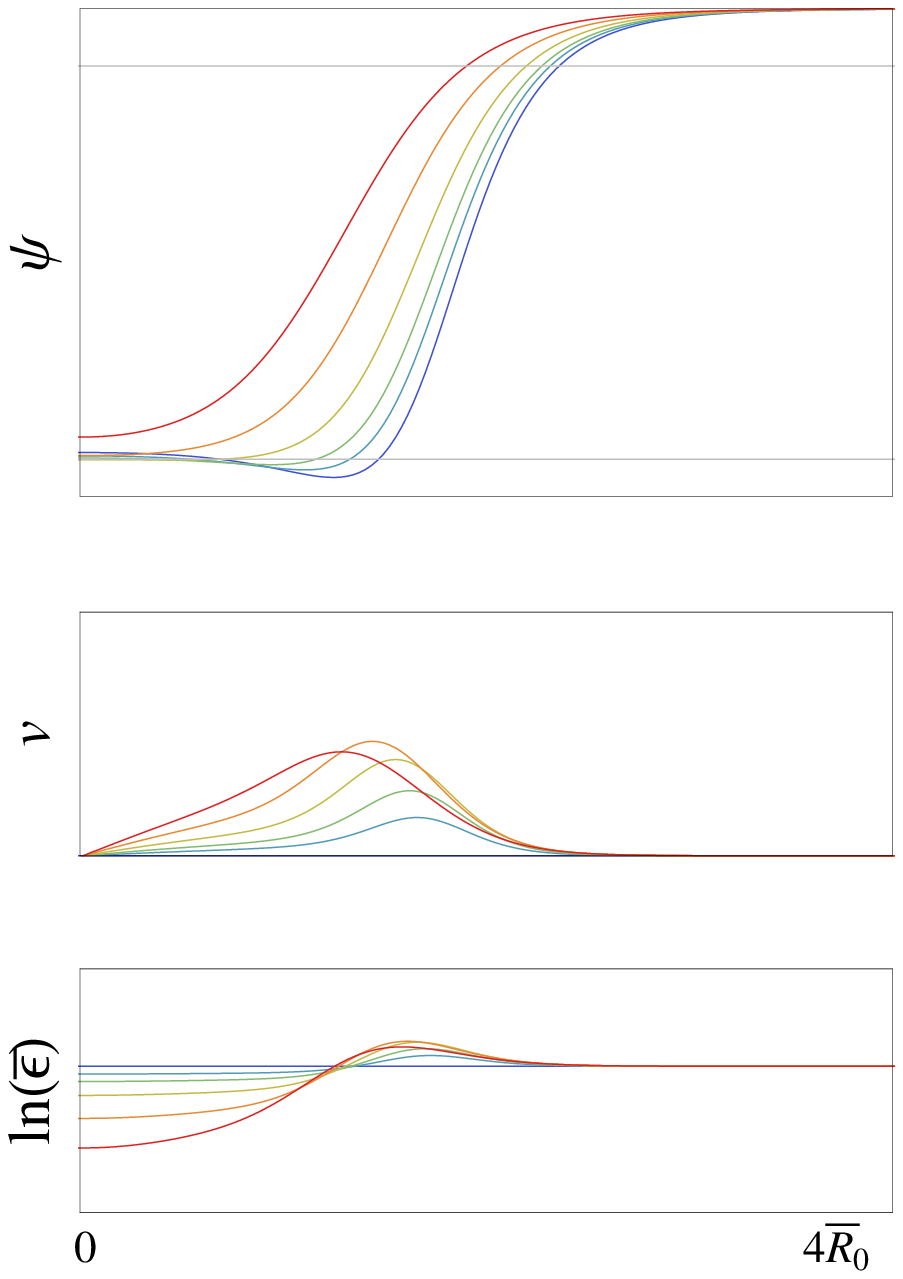}
      & \includegraphics[width=0.29\textwidth]{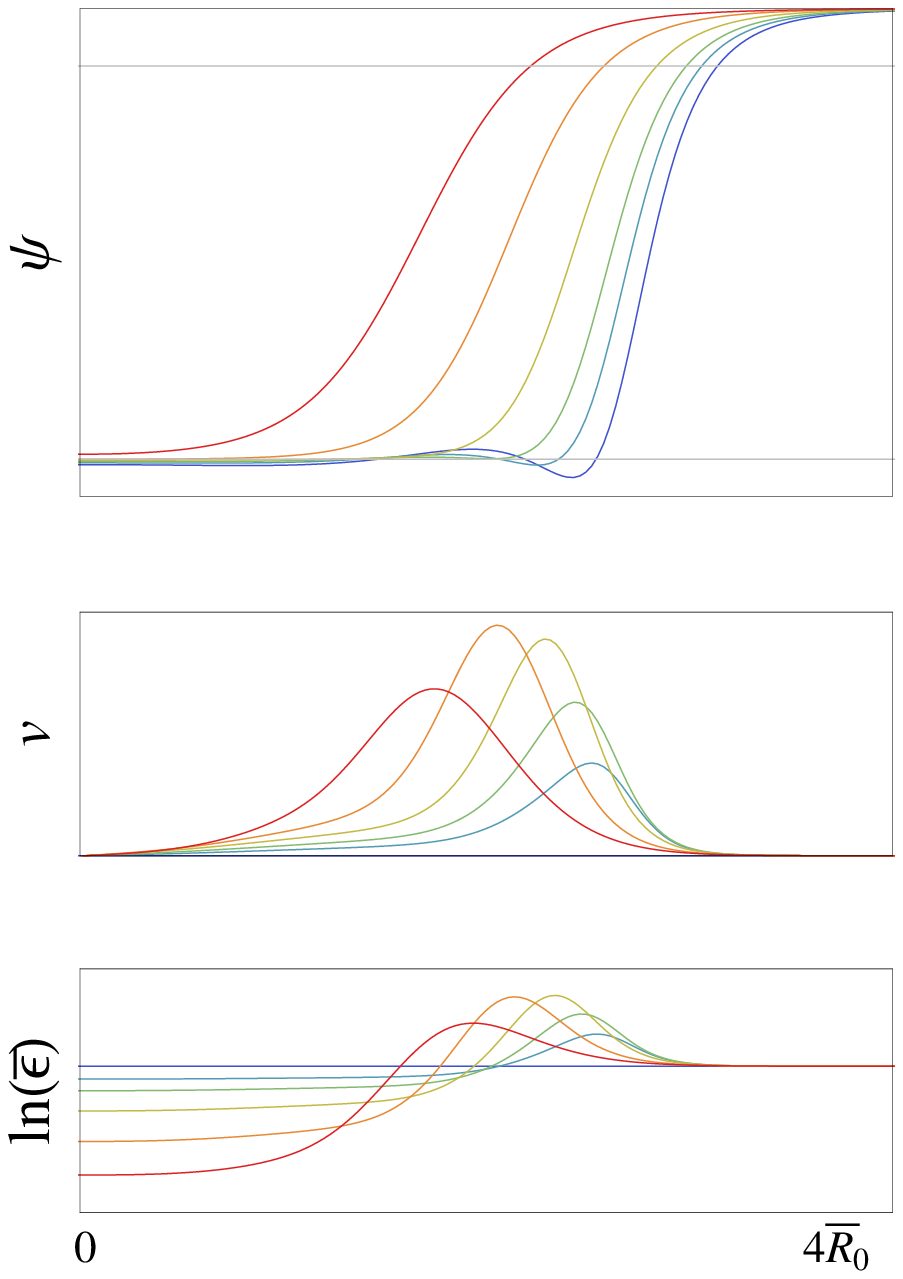}

  \end{tabular}
  \caption{The field profile, $\psi$, radial fluid velocity, $v$, and energy density profile, $\ln(\bar{\epsilon})$, for various coupling strengths (identified to the left of each set of plots) and potential differences: (a) Case I ($\alpha=0.96$), (b) Case II ($\alpha=0.65$) and (c) Case III ($\alpha = 0.45$). Grey horizontal lines in the field profile plots represent field values associated with the local extrema (the upper line in each plot is the maximum of the potential and the lower line is the true vacuum).  Horizontally these plots correspond to times when the radius of an uncoupled bubble reaches $\bar{R} = 0$, $\bar{R}=1.5\bar{R}_0$, and $\bar{R} = 2\bar{R}_0$.}
  \label{figevtab}
\end{figure*}

We allow the bubble to evolve until the radius of the bubble, $\bar{R}$, is at least twice its initial radius, $\bar{R}_f \gsim 2\bar{R}_0$.  First, we will describe the evolution of the bubbles as well as the behavior of the coupled fluid.  In Fig.~\ref{figevtab} we compare the field profile, the energy density and the radial velocity of the fluid for a one-dimensional slice through the center of the bubble as a function of the radius of the bubble, $\bar{r}$.  We plot three different time slices, chosen in each case to be the times when an uncoupled bubble reaches $\bar{R} = \left\{\bar{R}_0,1.5\bar{R}_0,2\bar{R}_0\right\}$, to show how the dynamical system changes as a function of coupling for each case.

To perform the actual integration, we use a memory-efficient variation of a second-order Runge-Kutta integrator.  This method, the `wedge' method, eliminates the need for multiple, full, copies of the grid at every time step.  In a standard second-order Runge-Kutta scheme, two full $N^3$ copies (where $N$ is the number of grid points along a side) are needed at every time step.  Here, we are able to reduce this need to be proportional to $N^2$ by building up a `wedge' of 2-dimensional slices that can be stored temporarily and replaced as we loop over the third dimension.  We are able to re-use memory allocated to area slices that are no longer needed, while preserving slices that are still required for adjacent evolution.  Once a `final' increment is calculated, the `final' copy of the grid can be overwritten.  This method can be used for explicit Runge-Kutta schemes of arbitrary order $n$, and given a grid of dimension $N^3$, will require only $\mathcal{O}(n^2N^2)$ additional memory.  This method can be implemented with OpenMP \cite{OpenMP} (or other parallel computing schemes) on the area of the grid, with the normal parallel computing idiosyncrasies.  We store simulation data using the HDF5 file format \cite{HDF5}.

Case I, the thin-wall limit, corresponds to $\alpha = 0.96$ .  This case is the most difficult to evolve numerically on a static grid, as the scale of the thickness of the bubble wall is much smaller than the size of the bubble and it is necessary to resolve both scales throughout the simulation.  For this simulation we use a cautiously high resolution $N^3 = 1024^3$, with grid cell spacing $dx = 0.005 \bar{R}_0$ and a time step of $dt = dx/10$.  Since the field gradients for the bubble are largest in this model, we see significant modification to the evolution of the bubble even for small couplings around $\xi \sim 10^{-2}$, and we examine up to $\xi = 0.2$.  Fig~\ref{figevtab} (a) shows the results of the simulations for $\alpha=0.96$.

In case II,  we examine a regime where we are no longer in the thin-wall limit, $\alpha = 0.65$.  We still use a high resolution ($N^3 = 768^3$) to ensure accuracy in results, although it is not necessary here to be as cautious as we were in case I.  For these simulations, we take $dx = 0.01 \bar{R}_0$ and $dt = dx/10$, and  $\xi$ varies between $0.06$ and $1.0$.  These simulations can be seen in Fig~\ref{figevtab} (b).

Finally, in case III we look at $\alpha = 0.45$ \cite{Hindmarsh:2013xza}.  These smaller values for $\alpha$ prove less computationally challenging since the scale of the thickness of the wall is approximately the same size as, or larger than, the radius of the bubble.  Since there is only one scale involved, lower resolution is possible.  We find very accurate results for $N^3 = 512^3$, $dx = 0.03 \bar{R}_0$, and $dt = dx/10$ and we vary $\xi$  from $0.1$ to $2.0$.  These simulations can be seen in Fig~\ref{figevtab} (c).

\begin{figure*}[hp]
    \begin{tabular}{ccc}
      $\alpha = 0.96$ & $\alpha = 0.65$ & $\alpha = 0.45$ \\
  \\ 
       \includegraphics[width=0.32\textwidth]{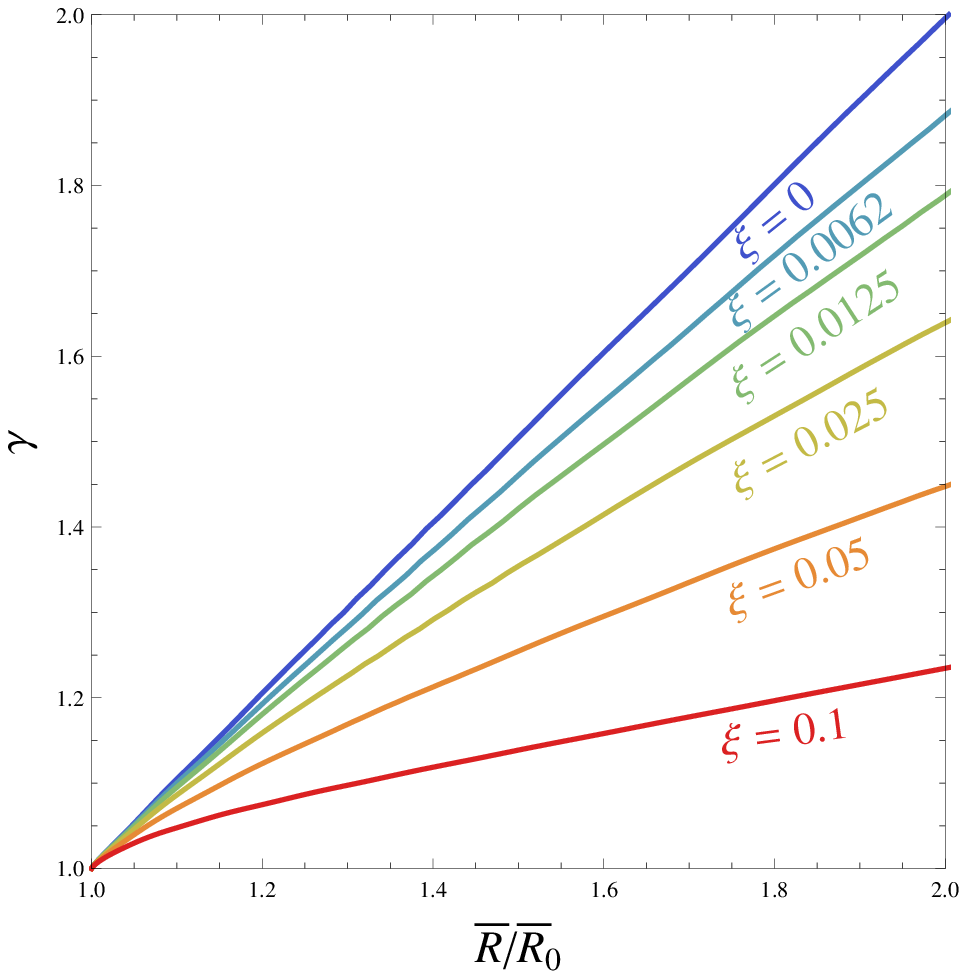}
        & \includegraphics[width=0.32\textwidth]{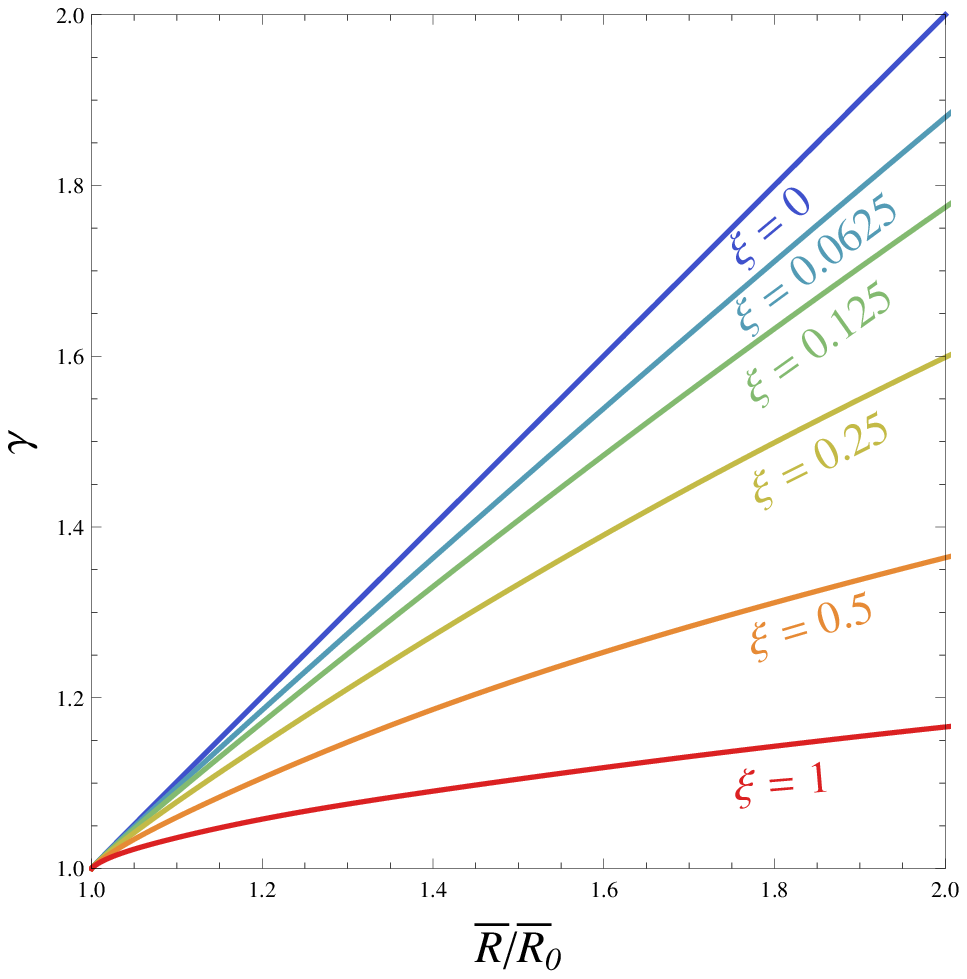}
        & \includegraphics[width=0.32\textwidth]{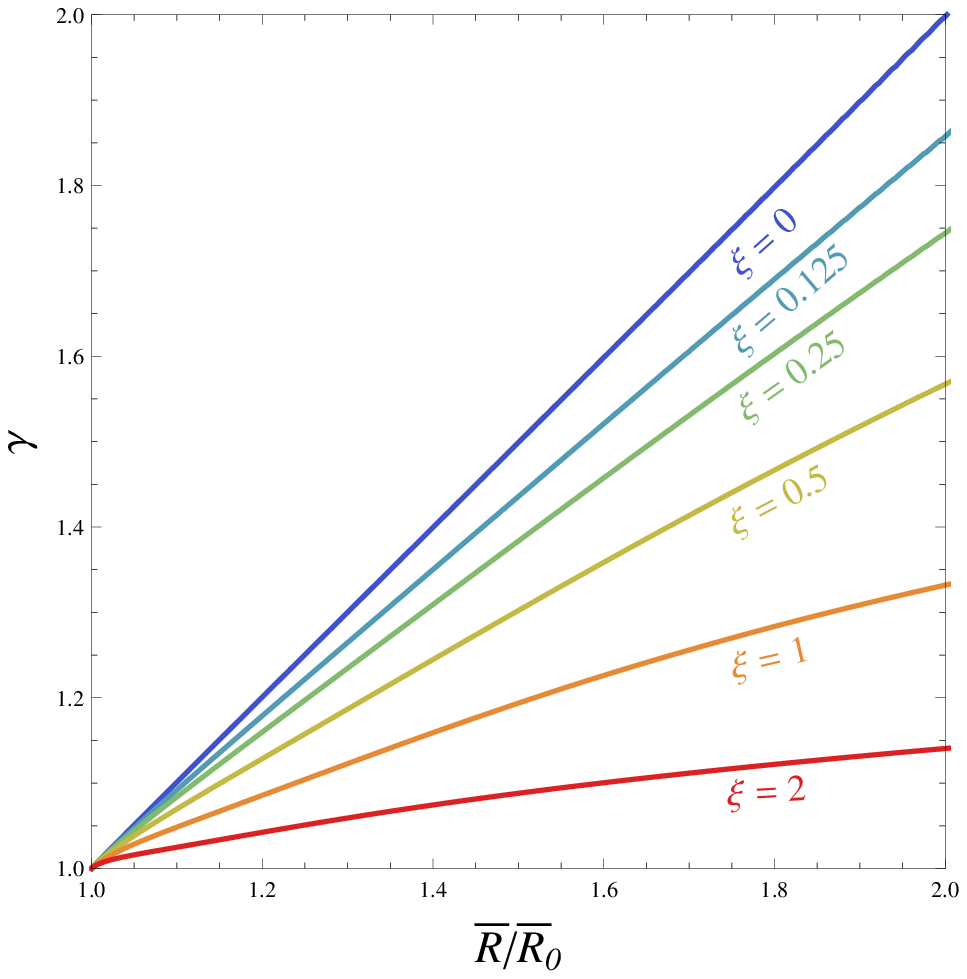} \\
  \\ \\ 
       \includegraphics[width=0.32\textwidth]{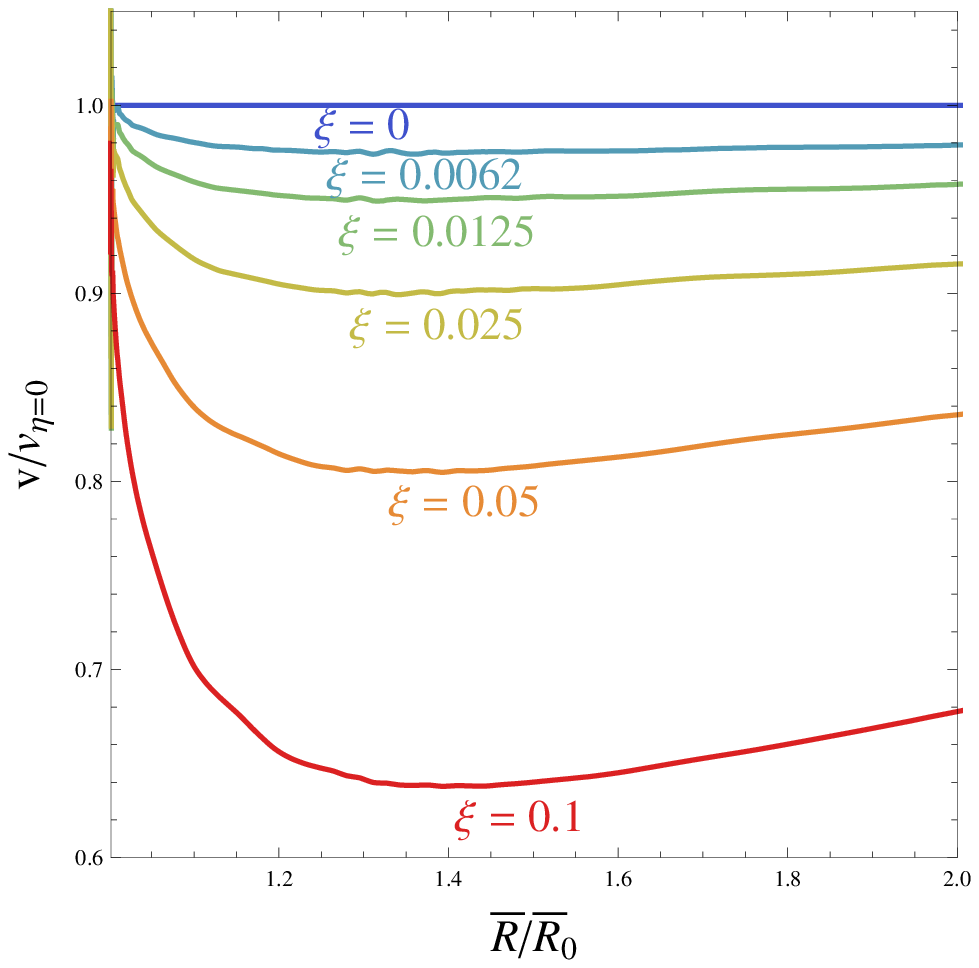}
        & \includegraphics[width=0.32\textwidth]{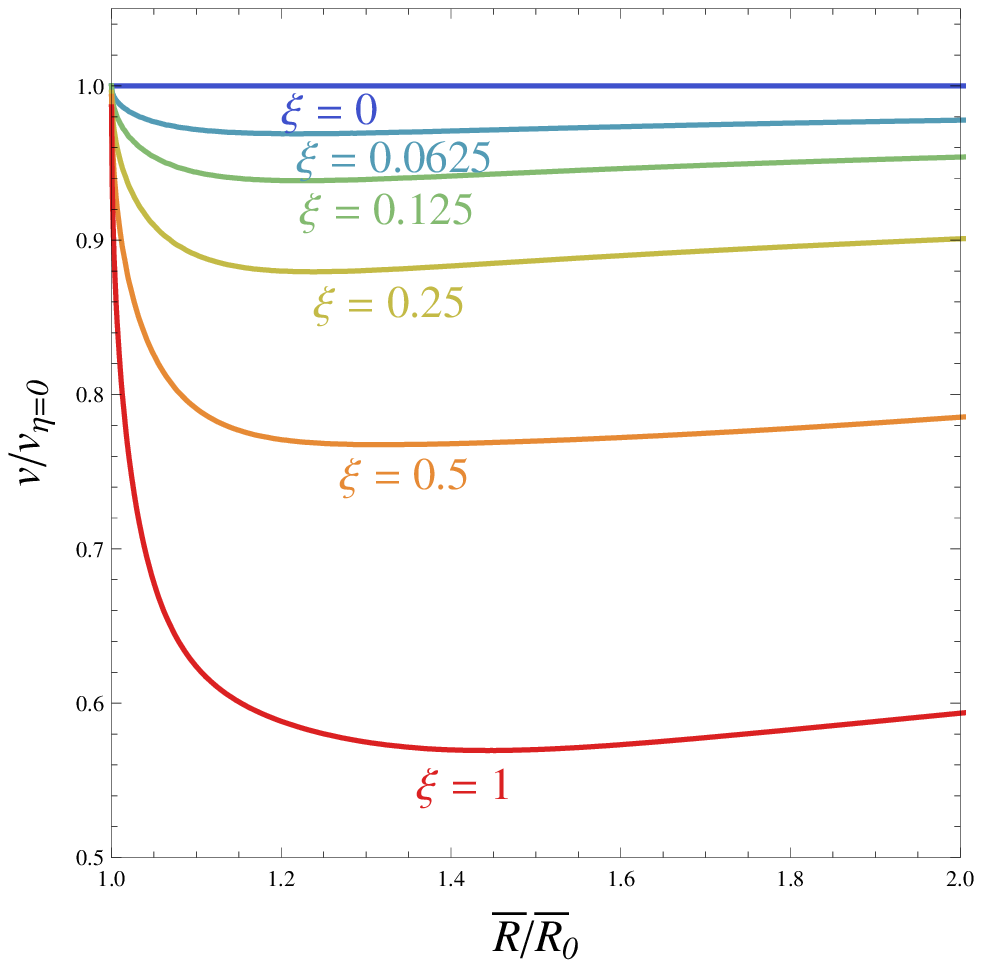}
        & \includegraphics[width=0.32\textwidth]{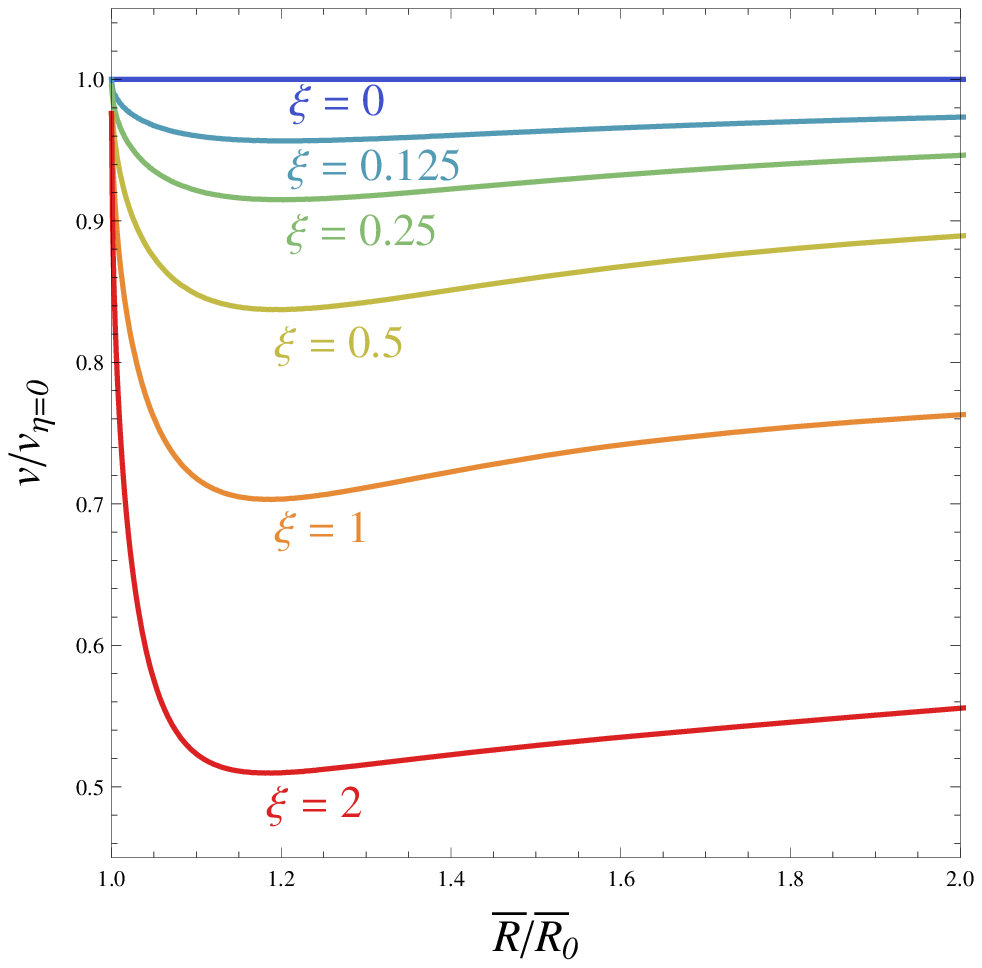} \\

    \end{tabular}
    \caption{Plots of the relativistic $\gamma$-factor of the bubble wall as a function of radius (top), and relative bubble wall velocities as a function of radius (bottom). }

    \label{figtesttab}
\end{figure*}

When the field is not coupled to the fluid, $\xi = 0$, the Euclidean bubble solution is $O(4)$-symmetric ($SO(3,1)$-symmetry in cosmic time) and we have an analytic expression for how a point on the initial bubble will evolve in time.  Specifically, as the bubble evolves, a point at radius $\bar{R}$ on the bubble wall obeys $\bar{R}^2+\tau^2=\bar{R}_0^2$ in the $\bar{R}-\tau$ plane (where $\tau = \imath \bar{t}$).  Equivalently, we expect that the relativistic $\gamma$-factor for the velocity of a point on the bubble wall should be $\bar{R}/\bar{R}_0$.  To confirm that our simulations are accurately capturing the dynamics of the bubble evolution, we track the bubble velocity as a function of how large the bubble has grown.  We calculate the velocity of the bubble wall by tracking a point on the bubble wall corresponding to the potential maximum at 
\begin{equation}
\psi = \frac{\sqrt{9-8 \alpha }-3}{2 \alpha}
\end{equation}
In Fig. \ref{figtesttab}, we plot this relationship, along with snapshots of the bubble and fluid evolution in Fig. \ref{figevtab}.

Fig.~\ref{figtesttab} shows that we are able to resolve the evolution of bubbles, even in the thin wall limits, and know that we can trust our simulations past the point where the bubble has grown by a factor of two.  We conservatively cut the simulation off at this point to make comparisons between the three cases of interest.  Since lower $\alpha$ simulations require much lower resolution, we would have the capacity to evolve bubbles for $\alpha\sim0.5$ for much longer while still trusting the simulations.  Further, Fig.~\ref{figtesttab} shows the parametric effect that the coupled fluid has on the velocity of the bubble.

The lower set of panes in Fig.~\ref{figtesttab} shows our main result--a striking effect on the late time velocity of the bubble wall.  Here we calculate the velocity of the bubble walls in the presence of a coupled fluid compared to the velocity of the bubble with $\xi = 0$.  As we increase the strength of the coupling, our bubbles have a (fairly constant) late-time velocity that is significantly different from the uncoupled case.  For some values of the coupling the velocity can be as low as $v=0.5c$, which significantly breaks the $SO(3,1)$ symmetry of the evolving bubble. 

We should note that these simulations are run for $\beta = 0.1$, the regime in which the energy density of the Universe is dominated by the fluid.  The values of the coupling strength $\xi$ are scaled logarithmically over a range of values chosen to illustrate the range of system behaviors.  When we move toward $\beta\rightarrow \infty$, we are less able to trust our temperature-independent assumptions; however, the software we present is capable of resolving dynamics in these regimes, so long as we retain a temperature-independent assumption.  

\section{Discussion}
\label{discussion}

We have presented a set of robust high-resolution numerical simulations of phase transitions in which the order-parameter that defines the phase transition is coupled to a dynamical, relativistic fluid.  We show that these simulations accurately track the behavior of the fluid and the field for a wide range of first-order phase transitions between $\alpha=0.45$ and $\alpha=0.96$, the thin-wall limit.  We have also shown that the velocities of the domain walls of these bubbles depend on the strength of the coupling between the fluid and field.  We show that even modest couplings cause the bubbles to slow down; breaking the $SO(3,1)$ symmetry associated with vacuum bubbles.  

We consider this work a first-step toward simulating the full dynamics of a first-order phase transition in the presence of a relativistic fluid, building on the numerical work already present in the literature and only the second such simulation to have been computed.

Since we trust the field and fluid evolution, we can use our code to simulate the generation of gravitational radiation in these models.  The software presented here can be easily `sewn in' to the software we used in \cite{Easther:2007vj}.  In the cases where $\alpha \approx 0.5$ we expect to be able to simulate many bubbles in a relatively large volume.  It will be a greater challenge in the cases where $\alpha \rightarrow 1$;  these simulations will likely not exceed $N^3 = 1024^3$ until new hardware is available.
 
Another open question probes the nature the coupling between the field and fluid.  The choice that we have made here represents a phenomenologically sound dissipative interaction, however it would be good to know whether new understandings of relativistic fluid actions \cite{Dubovsky:2011sj,Dubovsky:2011sk} can provide a  theoretically motivated interaction term.  Further it would be necessary to tie the effective parameters of this model, $\xi$ and the equation-of-state, $w$, to the parameters of the relevant fundamental particle physics. 

\section{Acknowledgments}  We would like to think Neil Barnaby for his inspiring thoughts on this topic as well as J. Tate Deskins and Ryan Darragh who conducted proof-of-concept simulations.  We are particularly thankful to Andrew Tolley, Glenn Starkman, and Shaden Smith for useful conversations.  JTG is supported by the National Science Foundation, PHY-1068080.  JM is supported by a Department of Education GAANN Fellowship.

The majority of our simulations were conducted on hardware provided by the National Science Foundation, the Research Corporation for Scientific Advancement and the Kenyon College Department of Physics.  Additional computing resources were provided by the High Performance Computing Cluster (HPCC) at Case Western Reserve University.

\end{document}